\PassOptionsToPackage{table,dvipsnames}{xcolor}

\documentclass[sigconf,nonacm]{acmart}

\usepackage{multirow}
\usepackage{makecell}
\usepackage{graphicx} 
\usepackage{svg}
\usepackage{epsfig}
\usepackage{hyperref}
\usepackage{subcaption} 

\AtBeginDocument{%
  }

\setcopyright{CC}
\copyrightyear{2025}
\acmYear{2025}
\acmConference[SIGIR 2025]{48th International ACM SIGIR Conference on Research and Development in Information Retrieval
}{July 13--18 2025}{Padua, Italy}
\acmISBN{978-1-4503-XXXX-X/2018/06}

\begin{document}

\title[Reproducibility of GEIA]{Information Leakage of Sentence Embeddings via Generative Embedding Inversion Attacks}

\author{Antonios Tragoudaras}
\authornote{Equal Contribution.}
\affiliation{%
  \institution{University of Amsterdam}
  \city{Amsterdam}
  \country{The Netherlands}
}
\email{antonios.tragoudaras@student.uva.nl}

\author{Theofanis Aslanidis}
\authornotemark[1]
\affiliation{%
  \institution{University of Amsterdam}
  \city{Amsterdam}
  \country{The Netherlands}
}
\email{theofanis.aslanidis@student.uva.nl}

\author{Emmanouil Georgios Lionis}
\authornotemark[1]
\affiliation{%
  \institution{University of Amsterdam}
  \city{Amsterdam}
  \country{The Netherlands}
}
\email{akis.lionis@student.uva.nl}

\author{Marina Orozco González}
\authornotemark[1]
\affiliation{%
  \institution{University of Amsterdam}
  \city{Amsterdam}
  \country{The Netherlands}
}
\email{marina.orozco.gonzalez@student.uva.nl}

\author{Panagiotis Eustratiadis}
\affiliation{%
  \institution{University of Amsterdam}
  \city{Amsterdam}
  \country{The Netherlands}
}
\email{p.efstratiadis@uva.nl}


\begin{abstract}
Text data are often encoded as dense vectors, known as embeddings, which capture semantic, syntactic, contextual, and domain-specific information. These embeddings, widely adopted in various applications, inherently contain rich information that may be susceptible to leakage under certain attacks. The GEIA framework highlights vulnerabilities in sentence embeddings, demonstrating that they can reveal the original sentences they represent. In this study, we reproduce GEIA's findings across various neural sentence embedding models.
Additionally, we contribute new analysis to examine whether these models leak sensitive information from their training datasets. We propose a simple yet effective method without any modification to the attacker's architecture proposed in GEIA. The key idea is to examine differences between log-likelihood for masked and original variants of data that sentence embedding models have been pre-trained on, calculated on the embedding space of the attacker. Our findings indicate that following our approach, an adversary party can recover meaningful sensitive information related to the pre-training knowledge of the popular models used for creating sentence embeddings, seriously undermining their security.
Our code is available on: \href{https://github.com/taslanidis/GEIA}{\textcolor{blue}{https://github.com/taslanidis/GEIA}}
\end{abstract}




\maketitle
\section{Introduction}
Language Models (LMs) have seen significant advancements in recent years, with substantial increases in their parameter spaces, leading to the development of Large Language Models (LLMs). These LLMs have been deployed in real-world applications, such as Chat-GPT~\cite{achiam2023gpt} and Gemini~\cite{team2023gemini}. However, deploying them in real-world scenarios necessitates ensuring the privacy and safety of users interacting with these systems~\cite{yao2024survey}. Most LMs, however, are primarily designed to excel at language tasks, often with less emphasis on addressing security concerns. It has been shown that LMs are prone to memorizing their training data, which can lead to the extraction of private information~\cite{carlini2019secret,thakkar2021understanding}. Therefore, it is crucial to evaluate the extent to which LMs are susceptible to leaking information encountered either during training or inference.

LMs transform input text into high-dimensional raw vectors, known as embeddings, which capture semantic, syntactic, contextual, and domain-specific information. These embeddings are further transformed into sentence embeddings~\cite{reimers2019sentence}. Notably, some words from the original sentence can be partially recovered from sentence embeddings, a type of attack referred to as \textit{embedding inversion}~\cite{song2020information, pan2020privacy}. Previous works on embedding inversion primarily focused on extracting attributable information. However, a potential malicious adversary might aim to reconstruct the original sentence, thereby extracting sensitive information.

In this paper we focus on the Generative Embedding Inversion Attack (GEIA) introduced by H. Li et al. in ``Sentence Embedding Leaks More Information than You Expect: Generative Embedding Inversion Attack to Recover the Whole Sentence''~\cite{GEIA}, henceforth referred to as the ``original paper''. They propose a generative model that takes sentence embeddings as input and attempts to recover the original sentences. This work aims to reproduce the original paper's results and validate its claims, which are summarized as follows:

\begin{itemize}
    \item[] \textbf{Claim 1:} \label{Claim_1} GEIA outperforms prior embedding inversion attacks in classification metrics.
    \item[] \textbf{Claim 2:} \label{Claim_2} GEIA can successfully recover sensitive information.
    \item[] \textbf{Claim 3:} \label{Claim_3} GEIA is adaptive and effective across various LM-based sentence embedding models, featuring various model architectures or training methods.
    \item[] \textbf{Claim 4:} \label{Claim_4} GEIA generates coherent and contextually similar sentences to the original inputs.
\end{itemize}

Further than reproducing the original experiments, we explore a new research question to further understand GEIA's capabilities, and contribute new experiments and analysis. Specifically, we aim to investigate whether GEIA can detect training data leakage. Concretely, we formulate the following research question:

\begin{itemize}
    \item[] \textbf{RQ 1:} \label{RQ_1} Do sentence embeddings leak sensitive information from the embedding model's training data?
\end{itemize}

In summary, we reproduce and confirm the original paper's claims, while expanding the scope of the research, demonstrating that GEIA can effectively uncover sensitive information that was masked in the input sentence of the sentence embedding model, but was part of its training set.

\section{Background and Related Work}
There has been a significant effort to assess and improve the privacy of LLMs. Various studies have been employed to showcase the vulnerability of LLMs, including prompt injection attacks \cite{yan2024backdooring, li2024evaluating, greshake2023not, perez2022ignore}, backdoor attacks \cite{qi2021mind, kurita2020weight, kandpal2023backdoor}, and jailbreak attacks \cite{deng2023jailbreaker, guo2024cold, li2023multi}. This reproducibility study focuses an embedding inversion attack, GEIA~\cite{GEIA}. For a comprehensive overview of different types of attacks that pose a threat to LLM privacy, we refer the avid reader to a more extensive survey~\cite{li2023privacy}.

\subsection{Adversarial Scenarios}
During an adversarial attack, the knowledge and accessibility of the victim model can be categorized based on the level of access. This is typically divided into three distinct settings:

\begin{enumerate}
    \item \textit{White-box}: In this setting, the attacker has complete knowledge of the victim model, including its architecture, parameters, and gradients.
    \item \textit{Gray-box} (or \textit{blue-box}): Here, the attacker has partial knowledge of the victim model, such as its architecture, or its probability estimates.
    \item \textit{Black-box}: In this setting, the attacker has no knowledge of the victim model's internal workings and can only access its output values.
\end{enumerate}
By these standards, GEIA is a gray-box attack, as it assumes access to the victim model's embeddings.

\subsection{Embedding Inversion Attacks}\label{subsec:framework_attacks}
GEIA \cite{GEIA} and previous approaches \cite{song2020information} build upon the use of contextualized embeddings produced by embedding models \cite{reimers2019sentence,gao2021simcse} that create contextualized sentence-level embeddings, useful for a variety of downstream tasks in natural language processing (e.g., text summarization, question answering). The purpose is to faithfully reconstruct the original sentence using an attacker model, which is typically a learned neural model. The attacker model must learn to reconstruct the original input sentence without any knowledge of the data seen by the victim embedding model.
The problem formulation can be described as follows: Given an arbitrary text $\textit{x}$, a victim embedding model with frozen parameters produces a contextualized sentence-level representation $\textit{f(x)}$ in some bottleneck latent space of the encoder's victim model (typically the model is an encoder or encoder-decoder). The attacker objective is to learn an inverse mapping \textit{$\Phi$} as to recover original text. This mapping is not one-to-one, as the victim model architecture utilized pooling operations in some intermediate layers, making the task of recovering the original text from the aggregated sentence representation non-trivial. The only assumption made is that the adversary has access to an auxiliary dataset $D_{\textit{aux}}$ from which the original sentences $\textit{x}$ can be sampled to query the victim model and produce sentence representations.
\subsection{Baseline Attacker Models}
Given the problem formulation, the authors of \cite{GEIA} compare their method with well-established baselines, introduced in \cite{song2020information}. Specifically, the two baselines differ in how they learn the inverse mapping \textit{$\Phi$}.
In the first setting, called Multi-Label Classification (MLC), a Feed-Forward Network (MLP~\cite{song2020information}) is trained using a standard binary cross-entropy loss to predict attributes of $\textit{x}$ over the entire vocabulary. This approach models each word independently, lacking the ability to capture the connection between words in a sequence.
The second method leverages a unidirectional Gated Recurrent Unit (GRU) \cite{chung2014empirical} and performs Multi-Set Prediction (MSP~\cite{welleck2018loss}), with the objective of maximizing the set of tokens not predicted at the current timestep\footnote{Referring to the temporal timestep of the GRU architecture.}.

\subsection{Limitations of Previous Work}
The authors highlight the shortcomings of the two previous approaches. They cannot model word order in a sentence, handle word repetitions, and are usually only able to invert non-meaningful words (e.g., stopwords). To tackle these limitation they propose their solution, which is the first true generative embedding inversion attack (GEIA) capable of reconstructing meaningful and coherent sentences from sentence embeddings, rather than only partially predicting words or sets of unordered words, as in the case of MLC and MSP, respectively.


\section{Methodology}

\begin{figure*}[t]
  \centering
  \includegraphics[width=0.9\textwidth]{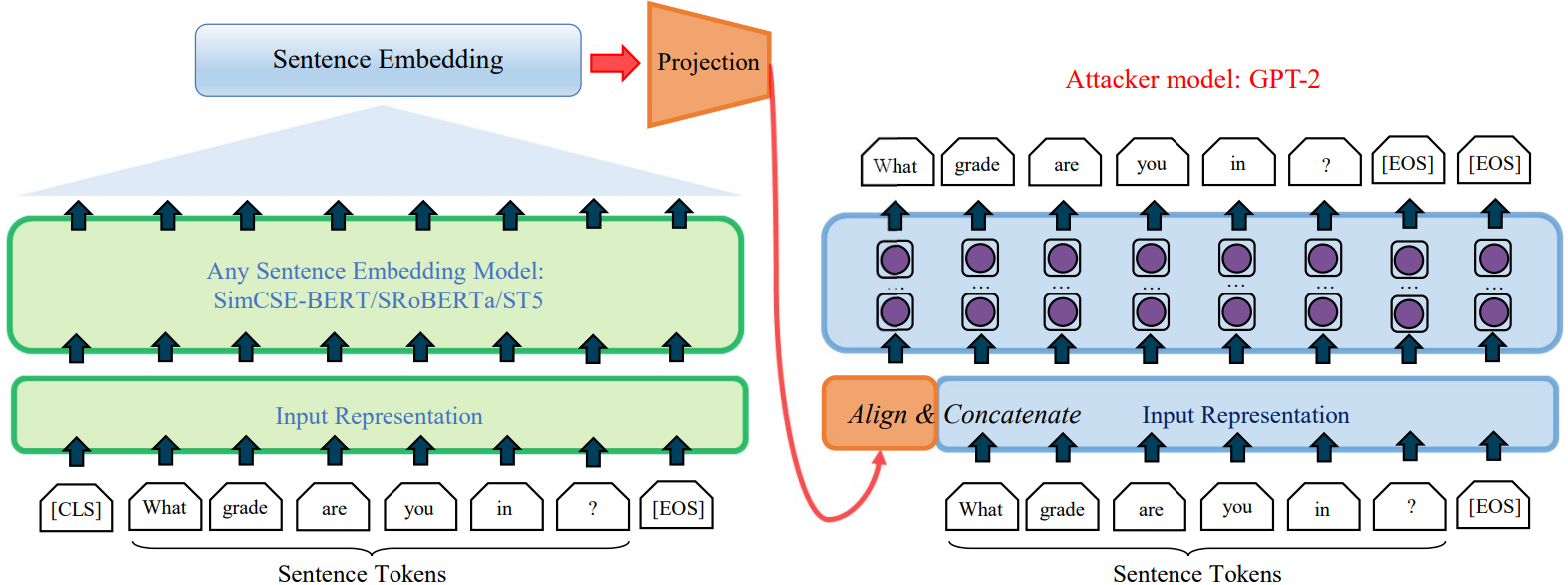}
  \caption{Architecture of GEIA. Illustration adopted from the original paper~\cite{GEIA}.}
  \label{fig:main-GEIA}
\end{figure*}

\subsection{GEIA}\label{subsection:GEIA}

GEIA builds upon the framework of Embedding Inversion Attacks, as delineated in Section \ref{subsec:framework_attacks}. Throughout this section, $f^{-1}$ is used interchangeably with $\Phi$, the notation introduced in earlier paragraphs, to refer to the inverse mapping. In contrast to previous approaches that reverse sentence embeddings, \cite{GEIA} is the first generative approach, achieving coherent and structured reconstruction. As shown in Fig. \ref{fig:main-GEIA}, the authors utilize a GPT-2 model, which will generate the input sentence given the provided sentence embedding. GEIA will utilize a randomly initialized decoder model to learn this inverse mapping $f^{-1}$ from a frozen pre-trained victim model. Since the output of the sentence embedding model (victim) is used as context for the attacker decoder, the embedding sizes must match. This is the reason the authors introduce a projection module, to align the sentence embeddings $f(x)$ in the embedding space of the decoder. However, this projection module is used only to align the embedding space dimensions and not to capture any extra semantic information.

\subsubsection*{\textbf{Training the Victim Model}} The sentence embedding model is pre-trained, and frozen throughout the whole training procedure. It's only the attacker model that is trying to learn the inverse mapping.

\subsubsection*{\textbf{Training the Attacker}} In order to train the attacker, we start from randomly initialized weights, in order to be comparable with baseline models. Also, the projection module $Align(f)$ is only used in cases of embedding size mismatch between the victim and the attacker. To achieve a coherent and semantically rich reconstruction, the decoder utilizes the sentence embedding $Align(f(x))$ and all previous context words $w_0,...,w_{i-1}$.

\begin{figure}[t]
    \centering
    \includegraphics[width=0.95\columnwidth]{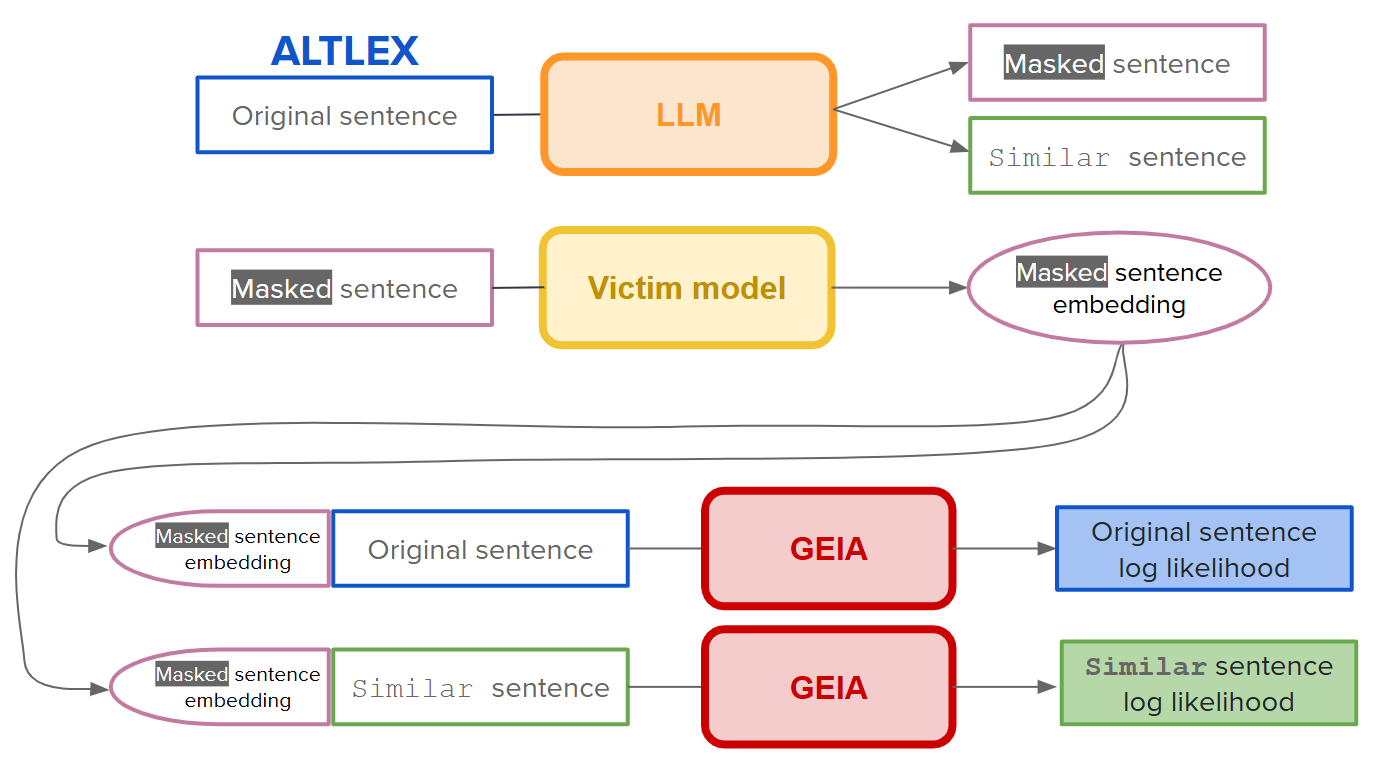}
    \caption{Pipeline overview of the method for assessing training data leakage.}
    \label{figure:first_extension_architecture}
\end{figure}

\subsection{Leakage from Training Data }\label{subsec:leakage_train_data}
In contrast to evaluating whether sentence embeddings leak information about the presented input text sampled from the datasets\footnote{The datasets used are introduced in Section \ref{subsec:Datasets}.} and used for evaluating the attacking performance, we propose that it is valuable to test the sensitivity of victim models in leaking information related to the data they were (pre)-trained on. We refer to this case as ``leakage from training data''. Figure \ref{figure:first_extension_architecture} provides an overview of the pipeline we have devised for assessing this potential vulnerability of the victim models.

For this purpose, we select a suitable dataset that the victim sentence embedding models have already been pre-trained on, namely AltLex \cite{hidey2016identifying}. This dataset contains diverse real-world facts extracted from Wikipedia articles.
In our pipeline, we employ  (decoder-only) language models (e.g GLM-4 \cite{glm2024chatglm}, and Llama-3 \cite{dubey2024llama} family model, details in Section \ref{few-show-reasoners-details}) as a few-shot reasoner to create masked and alternative variants of sentences sampled from AltLex. Specifically, the objective of the model is to obfuscate sensitive information from the original sentences, which can include names, locations, organizations, and any other named entities deemed important by the LLM reasoner. The model produces a masked sentence by replacing named entities in the original text with special tokens. Simultaneously, it generates an alternative version of the original sentence by substituting relevant words in place of the special tokens used to produce the masked sentence. The instruction template provided to the reasoners for performing this task using in-context learning is presented in Section \ref{subsection:prompts}. Examples of the masked and alternative/similar sentences are provided in Table \ref{tab:text_comparison}.

The masked sentences are then fed to the victim model, which creates a sentence embedding. This intermediate representation is pre-appended to both the original sentence and the alternative/similar sentence (the latter being generated by the reasoner). These two distinct representations are then fed to an attacker model\footnote{We use the learned weights for the GPT-2 model we trained from scratch on the PersonaChat dataset for reproducibility study, Section \ref{subsection:GEIA}.} that we have trained for the reproducibility study of GEIA. The attacker model projects these two hybrid representations into separate embedding spaces, for which we can calculate two associated likelihoods. Our assumption is that if the log-likelihood corresponding to the original sentence is consistently higher than the one corresponding to a similar sentence, it is intuitive to conclude that the victim model leaks information about the training data.
For our assumption to be sound, it is important to ensure that the leakage is isolated to the victim model, meaning that sensitive information is not broadcasted from the parametric memory of the attacker. To achieve this, the attacker has no knowledge of the data on which the victim model has been trained; it is only trained on a different auxiliary dataset, fully aligning with how the authors of GEIA trained their attacker. Furthermore, to ensure leakage isolation from the embeddings, we calculate the generation log-likelihood of the masked tokens from the attacker in two cases, with and without pre-appending the sentence embeddings $f(x)$ to the original and similar sentences given to the attacker.

\begin{table*}[tb]
\centering
\caption{Comparison of original, masked, and alternative text produced from GLM-4-9B-Chat on sampled sentences from AltLex.}

\begin{tabular}{@{}p{5cm}p{5cm}p{6cm}@{}}
\toprule
\textbf{Original} & \textbf{Masked} & \textbf{Alternative} \\
\midrule
Rommel was a commander of the F\"uhrerbegleithauptquartier (F\"uhrer escort headquarters) during the Poland campaign, often moving up close to the front in the F\"uhrersonderzug and seeing much of Hitler. & \textless PERSON\textgreater was a commander of the \textless ORGANIZATION\textgreater during the \textless LOCATION\textgreater campaign, often moving up close to the front in the \textless ENTITY\textgreater and seeing much of \textless PERSON\textgreater. & General von Kluge was a commander of the Supreme Command headquarters during the Warsaw campaign, often moving up close to the front in the High Command train and seeing much of the F\"uhrer. \\
\midrule
Rommel was born in Heidenheim, Germany, 45 kilometers (28 mi) from Ulm, in the Kingdom of W\"urttemberg which was then part of the German Empire, on November 15, 1891. & \textless PERSON\textgreater  was born in \textless LOCATION\textgreater, Germany, 45 kilometers (28 mi) from \textless LOCATION\textgreater, in the Kingdom of \textless ENTITY\textgreater, which was then part of the \textless ENTITY\textgreater, on November 15, 1891. & Karl was born in Munich, Germany, 45 kilometers (28 mi) from Augsburg, in the Kingdom of Bavaria, which was then part of the Austro-Hungarian Empire, on November 15, 1891. \\
\bottomrule
\end{tabular}
\label{tab:text_comparison}
\end{table*}

\section{Experimental Setup}

\subsection{Datasets}\label{subsec:Datasets}
\subsubsection*{\textbf{Dataset for Reproducibility}}
To ensure the reproducibility of results, we utilize the same datasets as the original authors, specifically two domain-specific datasets:
\begin{itemize}
    \item \textbf{PersonaChat (PC)}~\cite{personachat}: A persona-based conversational dataset comprising both real and synthetic personas.
    \item \textbf{QNLI}~\cite{wang-etal-2018-glue}: A dataset derived from Wikipedia articles, characterized by domain-specific vocabulary and syntax.
\end{itemize}

\subsubsection*{\textbf{Extended Dataset}}
For our extended experiments, we make use of the previously mentioned datasets along with additional ones to broaden the scope of our analysis:
\begin{itemize}
    \item \textbf{PersonaChat (PC)}~\cite{personachat}: This dataset was utilized in our training data leakage experiment, for training the attacker.
    \item \textbf{Altlex}~\cite{hidey2016identifying}: A dataset derived from Wikipedia articles and processed using parallel corpora to identify paraphrased pairs. It consists of parallel text segments, making it suitable for generating sentence perturbations and masked embeddings. Altlex also formed part of the training set for the sentence embedding models used.
\end{itemize}

\begin{table}[ht]
\centering
\caption{Statistics of datasets.}
\label{tab:dataset_stats}
\begin{tabular}{|l|c|c|c|}
\hline
\textbf{Stat Type} &  \textbf{\makecell{Persona\\Chat\\(PC)}} & \textbf{QNLI} &
\textbf{Altlex} \\
\hline
Task                  & Dialog  & NLI & \makecell{Causal\\connection}  \\ \hline
Domain                & Chit-chat & Wikipedia & Wikipedia  \\ \hline
\makecell[l]{Percentage\\Used}  & 100\% & 100\% & 10\% \\ \hline
Sentences             & 162,064  & 220,412 & 9,851  \\ \hline
\makecell[l]{Train:dev:test\\ split ratio} & 82:9:9 & 95:0:5  & 0:0:100 \\ \hline
\makecell[l]{Unique named\\ entities} & 1,425 & 46,567 & 12,000  \\ \hline
\makecell[l]{Avg. sentence\\ length}  & 11.71 & 18.25 & 21.38  \\ \hline
\end{tabular}
\end{table}

\subsection{Prompts}\label{subsection:prompts}
 In our extension (Section \ref{subsec:leakage_train_data}), we utilize fine-tuned instruction-based LLMs, as few-shot reasoners for the masking and alternative sentence tasks. The specific prompts we used for the instruction-based LLMs can be found in Appendix \ref{app:promptimg_schema}.

\subsection{Models}

\subsubsection*{\textbf{Victim Models}} As our scope is to reproduce faithfully the results of the original paper \cite{GEIA}, we consider the exact same victim model to produce sentence embeddings. Specifically, the models we used along with their pre-trained checkpoints are the following:\begin{itemize}
    \item Sentence-RoBERTa (SRoBERTa): \cite{reimers2019sentence}, \href{https://huggingface.co/sentence-transformers/all-roberta-large-v1}{\textcolor{blue}{weight-checkpoints}}.
    \item Sentence-T5 (ST5): \cite{ni2021sentence}, \href{https://huggingface.co/sentence-transformers/sentence-t5-large}{\textcolor{blue}{weight-checkpoints}}.
    \item MPNet: \cite{song2020mpnet}, \href{https://huggingface.co/sentence-transformers/all-mpnet-base-v1}{\textcolor{blue}{weight-checkpoints}}.
    \item SimCSE: \cite{gao2021simcse}, \href{https://huggingface.co/princeton-nlp/sup-simcse-bert-large-uncased}{\textcolor{blue}{SimCSE-BERT weight checkpoints}}, \href{princeton-nlp/unsup-simcse-roberta-large}{\textcolor{blue}{SimCSE-RoBERTa weight checkpoints}}.
\end{itemize}

\subsubsection*{\textbf{GEIA Model}} A random initialized GPT-2 medium\footnote{The model configuration for the GPT-2 medium model follows the structure of \href{https://huggingface.co/microsoft/DialoGPT-medium}{\textcolor{blue}{DialoGPT-medium}}.} decoder-only model serves as the attacker in the GEIA pipeline.

\subsubsection*{\textbf{LLM reasoners used in Leakage from Training Data}}\label{few-show-reasoners-details} Here we provide relevant information related to the reasoners used for creating the masked and alternative variants for sentences sampled from AltLex \cite{hidey2016identifying}, as described in subsection \ref{subsec:leakage_train_data}. We opted for the \href{https://huggingface.co/THUDM/glm-4-9b-chat/blob/main/README_en.md}{\textcolor{blue}{GLM-4-9b-chat}} from the GLM-4 family models \cite{glm2024chatglm}, the human preference-aligned version of the orginal \href{https://huggingface.co/THUDM/glm-4-9b/blob/main/README_en.md}{\textcolor{blue}{GLM-4-9B}} with enhanced instruction following capabilities needed for carrying out the masking task. One should note that, this is an important intermediate step in our pipeline as to determine if the victim models are prone to leak sensitive information related to the data have been trained on. The GLM-4-9B achieves similar performance to closed-soure state of the art LLMs, like Claude-3 \cite{claude}, GPT-4 \cite{achiam2023gpt}, Gemini 1.5 Pro \cite{team2023gemini}, while outperforming Llama 3 family models \cite{dubey2024llama} across various benchmarks used for assessing the model capabilities in semantics, mathematics, reasoning, code, and knowledge \cite{hendrycks2020measuring, cobbe2021training, chen2021evaluating, rein2023gpqa, wei2022chain}. The model has a context window of 128k tokens and has the expressive power to carry out the masking and similar sentence task adequately as demonstrated in Table \ref{tab:text_comparison}. All the latter reason along with its popularity in HuggingFace and similar parameter's model count to its counterpart Llama-based model, make its an excellent candidate for the task at hand. To make sure that our proposed pipeline (Fig.\ref{figure:first_extension_architecture}) is insensitive to the LLM selected for carrying out the masking and similar sentence derivation task we still decide to make use \href{https://huggingface.co/meta-llama/Llama-3.1-8B-Instruct}{\textcolor{blue}{Llama-3.1-8B-Instruct}} model. These two LLM reasoners we refer to them as GLM-4 and Llama3.1, respectively, for simplicity (Table \ref{tab:extension_results_1}).

\subsection{Hyperparameters}
We follow the exact same hyper-parameter choices in our study as to faithfully and fairly assess the reproducibility of the methods proposed in GEIA \cite{GEIA}.

\subsubsection*{\textbf{Baseline attacker models}} For the MLC task the authors of GEIA had carefully tuned the binary threshold of MLP with a grid-search algorithm on the validation split per victim model, and have all their results reported for the best discovered thresholds. Since they also share this threshold in their paper, we make direct use of them and report also our reproducibility result on the classification task (See Section \ref{subsec:repro_results}. The best threshold per victim model is depicted in Table \ref{tab:reproduced_results_1}). The MLC attacker setting is based on a single Feed-Forward network, while the MSP one leverage a uni-directional GRU of ten timesteps.

\subsubsection*{\textbf{Details for training GEIA}} Both the baseline and GEIA attackers are trained for ten epochs with a batch size of 64 samples. The Adam optimizer \cite{kingma2014adam} was used with a learning rate of $3e-4$, for minimizing the different objectives of each attacker.

\subsubsection*{\textbf{Tokenizer}} Both the baselines and GEIA leverage the byte-pair-encoding tokenizer \cite{sennrich2015neural}. Essentially this is very similar to the GPT-2 tokenizer, and the one we use is called \href{https://huggingface.co/microsoft/DialoGPT-medium}{\textcolor{blue}{DialoGPT-medium}}. Note that decoding uses beam search with beam size of five.

\subsubsection*{\textbf{LLM reasoners in Leakage from Training Data}}
For both GLM-4 and Llama-3.1, we use an inference batch size of 128 samples and set the generation hyperparameters to $top\_k=1$ and $max\_length=2500$. These settings, recommended by the authors of GLM-4, were found to yield the best performance on the masking task. Applying the same settings to Llama-3 also produced consistent results.

\subsection{Implementation Details}

All reproducibility experiments were performed with a single \textit{ NVIDIA A100-SXM4-40GB} GPU in a cluster environment, while for the masking task, as described in Section \ref{subsec:leakage_train_data} and Figure \ref{figure:first_extension_architecture}, we have leveraged the more powerful and latest GPU chip \textit{NVIDIA-H100-96GB}.

\section{Results and Analysis}

\begin{table*}[ht]
\centering

\caption{Evaluation of embedding inversion techniques, i.e., multi-label classification (MLC), multi-set prediction (MSP), and generative embedding inversion (GEIA), on the PersonaChat and QNLI datasets.  The token-level micro-averaged precision, recall, and F1 are reported, measured in \%. Reproduced results are shown in black, and the original results are in orange.}
\label{tab:reproduced_results_1}
\begin{tabular}{|c|c|c|c|c|c|c|c|c|c|c|c|}
\hline
\multirow{2}{*}{\textbf{Data}} & \multirow{2}{*}{\textbf{Victim Model}}  & \multicolumn{4}{c|}{\textbf{MLC}} & \multicolumn{3}{c|}{\textbf{MSP}} & \multicolumn{3}{c|}{\textbf{GEIA}} \\ \cline{3-12}
   &  & \textbf{Threshold} & \textbf{Pre} & \textbf{Rec} & \textbf{F1} & \textbf{Pre} & \textbf{Rec} & \textbf{F1} & \textbf{Pre} & \textbf{Rec} & \textbf{F1} \\ \hline

\multirow{5}{*}{PC}
& SRoBERTa      & 0.20 & \makecell{{78.38} \\ \textcolor{Bittersweet}{33.42}} & \makecell{12.38 \\ \textcolor{Bittersweet}{26.79}} & \makecell{21.38 \\ \textcolor{Bittersweet}{29.74}} & \makecell{56.13 \\ \textcolor{Bittersweet}{43.49}} & \makecell{45.33 \\ \textcolor{Bittersweet}{38.12}} & \makecell{50.16 \\ \textcolor{Bittersweet}{40.59}} & \makecell{56.64 \\ \textcolor{Bittersweet}{58.41}} & \makecell{{48.41} \\ \textcolor{Bittersweet}{48.91}} & \makecell{{52.78} \\ \textcolor{Bittersweet}{53.24}} \\ \cline{2-12}

& SimCSE-BERT   & 0.50 & \makecell{21.80 \\ \textcolor{Bittersweet}{24.77}} & \makecell{24.12 \\ \textcolor{Bittersweet}{21.36}} & \makecell{22.91 \\ \textcolor{Bittersweet}{22.94}} & \makecell{61.71 \\ \textcolor{Bittersweet}{42.23}} & \makecell{49.84 \\ \textcolor{Bittersweet}{37.10}} & \makecell{55.14 \\ \textcolor{Bittersweet}{39.50}} & \makecell{{66.53} \\ \textcolor{Bittersweet}{66.95}} & \makecell{{60.23} \\ \textcolor{Bittersweet}{59.69}} & \makecell{{63.22} \\ \textcolor{Bittersweet}{63.11}} \\ \cline{2-12}

& SimCSE-RoBERTa & 0.50 & \makecell{{84.28} \\ \textcolor{Bittersweet}{54.58}} & \makecell{33.65 \\ \textcolor{Bittersweet}{28.15}} & \makecell{48.10 \\ \textcolor{Bittersweet}{37.14}} & \makecell{63.91 \\ \textcolor{Bittersweet}{38.79}} & \makecell{51.62 \\ \textcolor{Bittersweet}{34.08}} & \makecell{57.11 \\ \textcolor{Bittersweet}{36.29}} & \makecell{64.20 \\ \textcolor{Bittersweet}{64.27}} & \makecell{{57.70} \\ \textcolor{Bittersweet}{56.66}} & \makecell{{60.79} \\ \textcolor{Bittersweet}{60.22}} \\ \cline{2-12}

& ST5 & 0.10 & \makecell{{76.86} \\ \textcolor{Bittersweet}{22.93}} & \makecell{12.41 \\ \textcolor{Bittersweet}{38.17}} & \makecell{21.38 \\ \textcolor{Bittersweet}{28.65}} & \makecell{60.53 \\ \textcolor{Bittersweet}{41.69}} & \makecell{48.89 \\ \textcolor{Bittersweet}{36.63}} & \makecell{54.09 \\ \textcolor{Bittersweet}{38.99}} & \makecell{66.63 \\ \textcolor{Bittersweet}{67.46}} & \makecell{{59.80} \\ \textcolor{Bittersweet}{58.26}} & \makecell{{63.04} \\ \textcolor{Bittersweet}{62.53}} \\ \cline{2-12}

& MPNet & 0.20 & \makecell{{78.70} \\ \textcolor{Bittersweet}{33.91}} & \makecell{12.37 \\ \textcolor{Bittersweet}{27.39}} & \makecell{21.38 \\ \textcolor{Bittersweet}{30.30}} & \makecell{57.57 \\ \textcolor{Bittersweet}{29.23}} & \makecell{46.50 \\ \textcolor{Bittersweet}{34.46}} & \makecell{51.44 \\ \textcolor{Bittersweet}{36.69}} & \makecell{60.75 \\ \textcolor{Bittersweet}{62.64}} & \makecell{{53.82} \\ \textcolor{Bittersweet}{53.51}} & \makecell{{57.07} \\ \textcolor{Bittersweet}{57.72}} \\ \hline \hline

\multirow{5}{*}{QNLI}
& SRoBERTa      & 0.20 & \makecell{{84.96} \\ \textcolor{Bittersweet}{44.73}} & \makecell{11.06 \\ \textcolor{Bittersweet}{19.68}} & \makecell{19.58 \\ \textcolor{Bittersweet}{27.33}} & \makecell{57.27 \\ \textcolor{Bittersweet}{47.42}} & \makecell{25.94 \\ \textcolor{Bittersweet}{22.47}} & \makecell{{35.71} \\ \textcolor{Bittersweet}{30.49}} & \makecell{41.50 \\ \textcolor{Bittersweet}{43.81}} & \makecell{{26.52} \\ \textcolor{Bittersweet}{27.19}} & \makecell{32.36 \\ \textcolor{Bittersweet}{33.56}} \\ \cline{2-12}

& SimCSE-BERT   & 0.60 & \makecell{15.93 \\ \textcolor{Bittersweet}{10.48}} & \makecell{18.48 \\ \textcolor{Bittersweet}{3.90}} & \makecell{17.11 \\ \textcolor{Bittersweet}{5.69}} & \makecell{{60.07} \\ \textcolor{Bittersweet}{46.43}} & \makecell{27.21 \\ \textcolor{Bittersweet}{22.00}} & \makecell{{37.45} \\ \textcolor{Bittersweet}{29.85}} & \makecell{47.30 \\ \textcolor{Bittersweet}{48.78}} & \makecell{{29.43} \\ \textcolor{Bittersweet}{29.49}} & \makecell{36.42 \\ \textcolor{Bittersweet}{36.76}} \\ \cline{2-12}

& SimCSE-RoBERTa & 0.75 & \makecell{4.44 \\ \textcolor{Bittersweet}{18.74}} & \makecell{21.40 \\ \textcolor{Bittersweet}{10.10}} & \makecell{7.36 \\ \textcolor{Bittersweet}{14.95}} & \makecell{{61.86} \\ \textcolor{Bittersweet}{52.57}} & \makecell{28.02 \\ \textcolor{Bittersweet}{24.90}} & \makecell{{38.57} \\ \textcolor{Bittersweet}{33.80}} & \makecell{48.05 \\ \textcolor{Bittersweet}{48.62}} & \makecell{{29.33} \\ \textcolor{Bittersweet}{29.26}} & \makecell{36.29 \\ \textcolor{Bittersweet}{36.53}} \\ \cline{2-12}

& ST5 & 0.20 & \makecell{{86.08} \\ \textcolor{Bittersweet}{42.26}} & \makecell{10.90 \\ \textcolor{Bittersweet}{19.83}} & \makecell{19.35 \\ \textcolor{Bittersweet}{27.00}} & \makecell{58.43 \\ \textcolor{Bittersweet}{48.50}} & \makecell{26.47 \\ \textcolor{Bittersweet}{22.98}} & \makecell{{36.43} \\ \textcolor{Bittersweet}{31.18}} & \makecell{46.00 \\ \textcolor{Bittersweet}{47.42}} & \makecell{{28.90} \\ \textcolor{Bittersweet}{28.43}} & \makecell{35.55 \\ \textcolor{Bittersweet}{35.55}} \\ \cline{2-12}

& MPNet & 0.45 & \makecell{{85.46} \\ \textcolor{Bittersweet}{53.25}} & \makecell{10.66 \\ \textcolor{Bittersweet}{10.29}} & \makecell{18.96 \\ \textcolor{Bittersweet}{17.24}} & \makecell{56.45 \\ \textcolor{Bittersweet}{47.18}} & \makecell{25.57 \\ \textcolor{Bittersweet}{22.35}} & \makecell{{35.19} \\ \textcolor{Bittersweet}{30.33}} & \makecell{42.47 \\ \textcolor{Bittersweet}{44.89}} & \makecell{{27.04} \\ \textcolor{Bittersweet}{27.74}} & \makecell{33.04 \\ \textcolor{Bittersweet}{34.29}} \\ \hline

\end{tabular}
\end{table*}

\subsection{Reproduced Results}\label{subsec:repro_results}
The reproduced results are presented in Tables \ref{tab:reproduced_results_1}, \ref{tab:reproduced_results_2}, and \ref{tab:reproduced_results_3}. On the one hand, we observe discrepancies in the baseline models performances between the original work and ours. In our case, the baseline models performed better than what the authors reported in their comparison tables. These differences may stem from randomization factors or other aspects that prevented us from fully replicating the original setup. On the other hand, GEIA results are nearly equal to the original paper, with discrepancies around 1-2\%, indicating that the proposed method is reproducible across different models.
Despite these discrepancies, the overall findings of the original authors are still valid, supporting the credibility of their claims and confirming the general trends observed in their results.

\subsubsection*{\hyperref[Claim_1]{\textbf{Claim 1:}} \textbf{GEIA outperforms previous embedding inversion attacks in classification metrics.}}

To evaluate the effectiveness of GEIA using classification metrics in conjunction with the baseline models, a clear distinction in performance emerges, particularly between non-generative models (e.g., MLC and MSP) and the generative model (GEIA), as summarized in Table~\ref{tab:reproduced_results_1}.

First, for classification tasks, both precision and recall are crucial, making the F1 score the primary metric of interest. On the Persona Chat dataset, GEIA achieves the highest F1 score and recall among all models. While the MLC model exhibits slightly higher precision compared to GEIA, its recall and F1 performance are significantly lower, indicating a weaker overall performance. Notably, GEIA outperforms the MSP model in F1 score by a margin of 2--10\%. For the QNLI dataset, MSP achieves the highest F1 score, with GEIA ranking second, trailing by only 2\%. However, GEIA demonstrates superior recall, whereas MSP excels in precision, leading to a higher F1 score for MSP.
In summary, while MSP marginally outperforms GEIA on QNLI by 2\% in F1 score, GEIA's superior recall and its strong performance on the Persona Chat dataset support \hyperref[Claim_1]{Claim 1}, albeit partially.

\begin{table*}[ht]
\centering
\caption{Performance of embedding inversion on stop word rate (SWR) and named entity recovery ratio (NERR), measured in \%. For SWR, we show the baseline values and the differences between baselines and SWRs from various attacks. A high NERR with a low SWR difference indicates effective token recovery, while a high SWR with low NERR suggests poor attack success despite good classification. Reproduced results are shown in black, and original results in orange.}
\label{tab:reproduced_results_2}
\begin{tabular}{|c|c|c|c|c|c|c|c|c|}
\hline
\multirow{2}{*}{\textbf{Data}} & \multirow{2}{*}{\textbf{Victim Model}}  & \multicolumn{4}{c|}{\textbf{SWR}} & \multicolumn{3}{c|}{\textbf{NERR}} \\
\cline{3-9}
 & &  \textbf{Test Set} &\textbf{MLC} & \textbf{MSP} & \textbf{GEIA} & \textbf{MLC} & \textbf{MSP} & \textbf{GEIA} \\
\hline
\multirow{5}{*}{PC}
& SRoBERTa & \makecell{52.74 \\ \textcolor{Bittersweet}{61.06}} & \makecell{{+04.00} \\ \textcolor{Bittersweet}{+38.80}} & \makecell{-20.00 \\ \textcolor{Bittersweet}{+25.69}} & \makecell{+07.31 \\ \textcolor{Bittersweet}{-05.01}} & \makecell{00.03 \\ \textcolor{Bittersweet}{00.05}} & \makecell{00.02 \\ \textcolor{Bittersweet}{00.05}} & \makecell{{26.00} \\ \textcolor{Bittersweet}{27.62}}  \\
\cline{2-9}
& SimCSE-BERT & \makecell{52.74 \\ \textcolor{Bittersweet}{61.06}} & \makecell{+22.44 \\ \textcolor{Bittersweet}{-20.50}} & \makecell{-16.72 \\ \textcolor{Bittersweet}{+27.58}} & \makecell{{+07.38} \\ \textcolor{Bittersweet}{-06.10}} & \makecell{00.05 \\ \textcolor{Bittersweet}{00.03}} & \makecell{00.40 \\ \textcolor{Bittersweet}{00.08}} & \makecell{{51.00} \\ \textcolor{Bittersweet}{55.57}}  \\
\cline{2-9}
& SimCSE-RoBERTa & \makecell{52.74 \\ \textcolor{Bittersweet}{61.06}} & \makecell{-17.69 \\ \textcolor{Bittersweet}{+00.52}} & \makecell{-19.22 \\ \textcolor{Bittersweet}{+34.49}} & \makecell{{+07.43} \\ \textcolor{Bittersweet}{-06.14}} & \makecell{00.34 \\ \textcolor{Bittersweet}{00.87}} & \makecell{01.70 \\ \textcolor{Bittersweet}{00.15}}  & \makecell{{48.00} \\ \textcolor{Bittersweet}{52.56}}\\
\cline{2-9}
& ST5 & \makecell{52.74 \\ \textcolor{Bittersweet}{61.06}} & \makecell{{+02.70} \\ \textcolor{Bittersweet}{+33.66}} & \makecell{-19.59 \\ \textcolor{Bittersweet}{-30.99}} & \makecell{+07.31 \\ \textcolor{Bittersweet}{-05.70}} & \makecell{00.02 \\ \textcolor{Bittersweet}{00.05}} & \makecell{00.02 \\ \textcolor{Bittersweet}{00.05}} & \makecell{43.00 \\ \textcolor{Bittersweet}{44.66}}  \\
\cline{2-9}
& MPNet & \makecell{52.74 \\ \textcolor{Bittersweet}{61.06}} & \makecell{{+04.24} \\ \textcolor{Bittersweet}{+38.83}} & \makecell{-20.21 \\ \textcolor{Bittersweet}{+30.54}} & \makecell{+07.04 \\ \textcolor{Bittersweet}{-05.31}} & \makecell{00.03 \\ \textcolor{Bittersweet}{00.05}} & \makecell{00.02 \\ \textcolor{Bittersweet}{00.05}} & \makecell{{31.16} \\ \textcolor{Bittersweet}{32.50}} \\
\hline
\hline
\multirow{5}{*}{QNLI}
& SRoBERTa & \makecell{31.18 \\ \textcolor{Bittersweet}{38.13}} & \makecell{-33.46 \\ \textcolor{Bittersweet}{+56.83}} & \makecell{-37.13 \\ \textcolor{Bittersweet}{+40.55}} & \makecell{{-04.26} \\ \textcolor{Bittersweet}{+05.14}} & \makecell{00.21 \\ \textcolor{Bittersweet}{01.06}} & \makecell{02.37 \\ \textcolor{Bittersweet}{02.12}} & \makecell{{12.48} \\ \textcolor{Bittersweet}{15.12}} \\
\cline{2-9}
& SimCSE-BERT & \makecell{31.18 \\ \textcolor{Bittersweet}{38.13}} & \makecell{+04.94 \\ \textcolor{Bittersweet}{-18.79}} & \makecell{-35.72 \\ \textcolor{Bittersweet}{+40.97}} & \makecell{{-03.79} \\ \textcolor{Bittersweet}{+04.04}} & \makecell{00.98 \\ \textcolor{Bittersweet}{00.10}} & \makecell{02.52 \\ \textcolor{Bittersweet}{1.84}} & \makecell{{15.15} \\ \textcolor{Bittersweet}{16.53}}  \\
\cline{2-9}
& SimCSE-RoBERTa & \makecell{31.18 \\ \textcolor{Bittersweet}{38.13}} & \makecell{+13.04 \\ \textcolor{Bittersweet}{-00.06}} & \makecell{-36.29 \\ \textcolor{Bittersweet}{+37.39}} & \makecell{{-03.69} \\ \textcolor{Bittersweet}{+03.65}} & \makecell{01.90 \\ \textcolor{Bittersweet}{00.82}} & \makecell{02.98 \\ \textcolor{Bittersweet}{02.50}} & \makecell{{16.60} \\ \textcolor{Bittersweet}{18.16}}  \\
\cline{2-9}
& ST5 & \makecell{31.18 \\ \textcolor{Bittersweet}{38.13}} & \makecell{-32.62 \\ \textcolor{Bittersweet}{+56.77}} & \makecell{-37.19 \\ \textcolor{Bittersweet}{+39.35}} & \makecell{{-03.08} \\ \textcolor{Bittersweet}{+04.45}} & \makecell{00.16 \\ \textcolor{Bittersweet}{01.06}} & \makecell{02.42 \\ \textcolor{Bittersweet}{02.09}} & \makecell{{14.98} \\ \textcolor{Bittersweet}{14.00}}  \\
\cline{2-9}
& MPNet & \makecell{31.18 \\ \textcolor{Bittersweet}{38.1
3}} & \makecell{-31.97 \\ \textcolor{Bittersweet}{+61.87}} & \makecell{-38.72 \\ \textcolor{Bittersweet}{+41.16}} & \makecell{{-03.50}\\ \textcolor{Bittersweet}{+04.31}} & \makecell{00.08 \\ \textcolor{Bittersweet}{00.70}} & \makecell{02.40 \\ \textcolor{Bittersweet}{01.97}} & \makecell{{12.14} \\ \textcolor{Bittersweet}{15.03}} \\
\hline
\end{tabular}
\end{table*}

\subsubsection*{\hyperref[Claim_2]{\textbf{Claim 2:}} \textbf{GEIA can successfully recover sensitive information}}

While precision, recall, and F1 score are critical metrics for evaluating the success of a classification attack, they do not capture the extent of sensitive information leakage. Table~\ref{tab:reproduced_results_2} presents the SWR (Sensitive Word Recall) and NERR (Normalized Entity Recall Rate) metrics, which measure the degree of sensitive information leakage.  It is important to note that the table highlights the difference between the ground truth SWR (i.e., the test set) and the model's SWR. The best model minimizes this difference, whether positive or negative. Additionally, the optimal model achieves the highest NERR value while maintaining a small SWR difference.

For the SWR metric, GEIA outperforms in most cases. However, on the Persona Chat (PC) dataset with the victim models SRoBERTa, ST5, and MPNet, MLC demonstrates better performance, with GEIA trailing by a margin of only 5\%.
In contrast, for the NERR metric, GEIA achieves the highest performance across both the PC and QNLI datasets, surpassing all other models. As expected, MLC exhibits the worst performance, given its classification-based nature. MSP, a set-word prediction model, performs moderately, achieving only a 3\% improvement on the QNLI dataset.As GEIA simultaneously achieves the best performance in both SWR and NERR, it demonstrates its capability to recover sensitive information, thereby supporting \hyperref[Claim_2]{Claim 2}.

\subsubsection*{\hyperref[Claim_3]{\textbf{Claim 3:}} \textbf{GEIA is adaptive and effective to various LM-based sentence embedding models, featuring various model architectures or training methods.}}

An effective adversarial attack should be capable of targeting different victim models without significant performance degradation. Tables~\ref{tab:reproduced_results_1} and \ref{tab:reproduced_results_2} demonstrate that GEIA maintains its effectiveness across various models. Specifically, as shown in Table~\ref{tab:reproduced_results_1}, on the Persona Chat (PC) dataset, the F1 score exhibits a deviation of approximately 10\% between models, while on the QNLI dataset, the deviation is only 4\%.

Furthermore, in Table~\ref{tab:reproduced_results_2}, the SWR metric shows a deviation of less than 1\% for both the PC and QNLI datasets. For the NERR metric, there is a larger deviation of around 25\% on the PC dataset, whereas on the QNLI dataset, the deviation drops to only 3\%. These consistent results across metrics and datasets demonstrate that the variation between victim models is generally minimal, thereby supporting \hyperref[Claim_3]{Claim 3}.

\subsubsection*{\hyperref[Claim_4]{\textbf{Claim 4:}} \textbf{GEIA generates coherent and contextually similar sentences as the original inputs.}} To evaluate the generation quality of GEIA, Table \ref{tab:reproduced_results_3} compares GEIA's performance across various generation quality metrics with the original results reported by the authors. Overall, we observe a high degree of similarity across most metrics, with discrepancies of up to 4\% between our reproduced results and the original findings. However, larger differences are noted for PPL and BLEU-2, with discrepancies ranging from 5\% to 17\%. We attribute these differences to the inherent stochasticity of the generation process, as the original authors did not provide specific seeds to control for variability. Additionally, for calculating PPL values, the original authors employed a fine-tuned GPT-2 model that we were unable to locate. As a result, we used a different GPT-2 model for our experiments. Despite these variations, for metrics where larger discrepancies are observed, our reproduced results outperform the original ones. This indicates that GEIA is capable of generating coherent and contextually appropriate sentences based on the input, further supporting \hyperref[Claim_4]{Claim 4}.

\begin{table*}[ht]
\centering
\caption{Evaluation of generation quality for generative embedding inversion attacks on victim and baseline embedding models. ES refers to embedding similarity and PPL to the perplexity of a GPT-2 model. Embedding similarity, ROUGE, and BLEU are reported in \%. The two GPT-2 models serve as baselines, generating sequences from the first input token with or without context. Reproduced results are shown in Black, while the original results are in orange.}
\label{tab:reproduced_results_3}
\begin{tabular}{|c|c|c|c|c|c|c|c|}
\hline
\multirow{2}{*}{\textbf{Model}} & \multirow{2}{*}{\textbf{PPL}} & \multirow{2}{*}{\textbf{ES}} & \multicolumn{2}{c|}{\textbf{ROUGE}} & \multicolumn{3}{c|}{\textbf{BLEU}}  \\
\cline{4-8}
& & & \textbf{ROUGE-1} & \textbf{ROUGE-L} & \textbf{BLEU-1} & \textbf{BLEU-2} & \textbf{BLEU-4}  \\ \hline
\makecell{SRoBERTa}            & \makecell{19.12 \\ \textcolor{Bittersweet}{4.99}} & \makecell{88.36 \\ \textcolor{Bittersweet}{88.07}} & \makecell{28.63 \\ \textcolor{Bittersweet}{59.54}} & \makecell{55.75 \\ \textcolor{Bittersweet}{56.04}} & \makecell{36.25 \\ \textcolor{Bittersweet}{35.47}} & \makecell{27.11 \\ \textcolor{Bittersweet}{20.37}} & \makecell{15.53 \\ \textcolor{Bittersweet}{15.66}} \\ \hline

\makecell{SimCSE-BERT}            & \makecell{17.40 \\ \textcolor{Bittersweet}{6.29}} & \makecell{91.86 \\ \textcolor{Bittersweet}{91.28}} & \makecell{69.42 \\ \textcolor{Bittersweet}{72.38}} & \makecell{62.87 \\ \textcolor{Bittersweet}{65.33}} & \makecell{44.28 \\ \textcolor{Bittersweet}{46.93}} & \makecell{34.00 \\ \textcolor{Bittersweet}{28.99}} & \makecell{20.20 \\ \textcolor{Bittersweet}{22.85}}  \\ \hline

\makecell{SimCSE-RoBERTa}            & \makecell{20.78 \\ \textcolor{Bittersweet}{5.98}} & \makecell{91.76 \\ \textcolor{Bittersweet}{91.33}} &  \makecell{71.87 \\ \textcolor{Bittersweet}{68.78}} & \makecell{66.03 \\ \textcolor{Bittersweet}{62.42}} & \makecell{47.31 \\ \textcolor{Bittersweet}{43.41}} & \makecell{37.13 \\ \textcolor{Bittersweet}{25.66}} & \makecell{22.94 \\ \textcolor{Bittersweet}{19.82}} \\ \hline

\makecell{ST5}            & \makecell{20.70 \\ \textcolor{Bittersweet}{5.90}} & \makecell{97.76 \\ \textcolor{Bittersweet}{91.47}} & \makecell{71.87 \\ \textcolor{Bittersweet}{70.72}} & \makecell{66.03 \\ \textcolor{Bittersweet}{65.45}} & \makecell{46.10 \\ \textcolor{Bittersweet}{44.52}} & \makecell{36.37 \\ \textcolor{Bittersweet}{27.83}} & \makecell{22.70 \\ \textcolor{Bittersweet}{21.99}}  \\ \hline

\makecell{MPNet}            & \makecell{20.12 \\ \textcolor{Bittersweet}{5.64}} & \makecell{89.35 \\ \textcolor{Bittersweet}{89.27}} &  \makecell{64.74 \\ \textcolor{Bittersweet}{65.08}} & \makecell{60.15 \\ \textcolor{Bittersweet}{60.39}} & \makecell{40.04 \\ \textcolor{Bittersweet}{40.04}} & \makecell{40.47 \\ \textcolor{Bittersweet}{23.83}} & \makecell{18.54 \\ \textcolor{Bittersweet}{18.54}} \\ \hline
\end{tabular}
\end{table*}

\subsection{Results on Leaking Training Data}\label{subsection:leakage_train_data}
\begin{table*}[htb]
\centering

\caption{Evaluation of the attacker's log likelihood distribution between generating original and similar sentences. The first part of the table consists of the mean probability over all tokens in the sentence, while the second part mean probability over the perturbed/masked tokens only. All results are statistically significant based on t-tests with p-values close to zero.}
\label{tab:extension_results_1}
\begin{tabular}{|c|c|p{0.1\linewidth}|p{0.1\linewidth}|p{0.1\linewidth}|p{0.1\linewidth}|p{0.1\linewidth}|p{0.1\linewidth}|}
\hline
\multirow{2}{*}{\textbf{LLM Reasoner}} & \multirow{2}{*}{\textbf{Victim Model}}  & \multicolumn{2}{c|}{\textbf{Original Likel. Distr. Mean}} & \multicolumn{2}{c|}{\textbf{Similar Likel. Distr. Mean}} & \multicolumn{2}{p{0.2\linewidth}|}{\textbf{Distr. Comparison Original vs Similar}} \\ \cline{3-8}
   &  & \textbf{w/ $f(x)$} & \textbf{w/o $f(x)$} & \textbf{w/ $f(x)$} & \textbf{w/o $f(x)$} & \textbf{w/ $f(x)$} & \textbf{w/o $f(x)$} \\ \hline

\multicolumn{8}{|c|}{\textbf{Mean log likelihood for the whole sentence}} \\
\hline

\multirow{2}{*}{GLM-4}

& SRoBERTa & -5.15 & -4.83 & -5.61 & -5.44 & +8.93\% & +12.63\% \\
& SimCSE-BERT & -5.56 & -4.91 & -5.96 & -5.36 & +7.19\% & +9.16\% \\

\hline

\multirow{2}{*}{Llama-3.1}
& SRoBERTa & -5.28 & -4.94 & -5.62 & -5.44 & +6.44\% & +10.12\% \\
& SimCSE-BERT & -5.71 & -5.03 & -6.00 & -5.37 & +5.08\% & +6.76\% \\

\hline

\multicolumn{8}{|c|}{\textbf{Mean log likelihood for the masked tokens only}} \\
\hline

\multirow{2}{*}{GLM-4}

& SRoBERTa & -10.4 & -11.05 & -13.33 & -11.26 & +28.17\% & +1.9\% \\
& SimCSE-BERT & -10.66 & -11.06 & -13.36 & -11.20 & +25.33\% & +1.27\% \\

\hline

\multirow{2}{*}{Llama-3.1}
& SRoBERTa & -9.99 & -11.08 & -13.36 & -11.04 & +33.73\% & -0.36\% \\
& SimCSE-BERT & -10.21 & -11.11 & -13.26 & -10.99 & +29.87\% & -1.08\% \\

\hline

\end{tabular}
\end{table*}

In this section, we evaluated whether sentence embeddings leak sensitive information from the training data. We will name three key components for this process: (1) the original sentence and the two outputs from the LLM, (2) the similar sentence, and (3) the masked sentence. Given the masked sentence, we create the sentence embeddings, and concatenate them with the original as one possible output for the attacker, as well as concatenate them with the similar as another possible output for the attacker. The assumption as stated previously, is that a consistently higher likelihood of the attacker reconstructing the original, rather than the similar, denotes the sentence embedding leaking some information on the embedding that is not there in the masked input text.

We tried to contain as much as possible, any parametric knowledge from the attacker impacting the probabilities of generating the masked tokens. We did this in two ways. First, by separating the domains of the input data each component was trained on. Furthermore, to further isolate the association of the sentence embedding with the likelihood of inverting the original tokens, we calculated the same likelihood distributions for the attacker with and without concatenating the sentence embedding $f(x)$. This will operate as a benchmark, to observe whether probabilities are higher when the attacker has access to the original $f(x)$, hinting to a leakage from this exact embedding.
In Table \ref{tab:extension_results_1}, we used a combination of different LLM reasoners, and victim models, to calculate the probabilities of generation of the masked original and similar sentence tokens, based on the sentence embedding $f(x)$. We also, calculated the probability of generating the masked tokens without any context $f(x)$ (purely generating the words $w_0,...,w_{i}$. Then, by comparing these two results, we can observe whether any leakage stems from the attacker model, or from the sentence embedding $f(x)$.

For instance, a $+30\%$ difference in distribution means, as seen in the most right columns of Table \ref{tab:extension_results_1}, means that it is 30\% more probable to generate the original tokens than it is to generate similar ones. Furthermore, we calculated these probabilities in two ways. The upper part of the table is by averaging the whole sentence logit probabilities of the attacker (decoder), while the bottom part, is by averaging only the masked/perturbed tokens log likelihood. Then given the mean likelihood for each sample, we have one likelihood distribution for the original tokens and one for the similar/perturbed tokens. Given these two likelihood distributions, we compared them by running t-tests, which concluded that all of our results have statistically significant differences under 95\% statistical power.

By observing the results in table \ref{tab:extension_results_1}, we can see that aggregating the probability over all tokens in the sentence, doesn't point to specific leakage, since the difference exists also when the attacker doesn't have access to the sentence embedding $f(x)$.

However, when aggregating only the masked tokens likelihood the conclusions are interesting. Given that without the sentence embedding $f(x)$ original and similar masked tokens have the same likelihood, when we introduce $f(x)$ to the attacker, then it's significantly more likely for the attacker to invert the original tokens that were masked, instead of similar ones. \textbf{This pattern hints to the conclusion that the masked sentence embeddings do carry significant sensitive information regarding the train set}, thus posing a considerable threat for embedding models, since attackers are able to identify exact sensitive information that the embedding model saw during training.\\
\textit{{\textbf{Qualitative analysis.}}} For the example of row 2 in Table \ref{tab:text_comparison}, when we don't use the sentence embedding of the masked text (column 2) the log-likelihood of generating the original tokens is -11.2, while generating the similar tokens is -10.14. On the other hand, when using the sentence embedding of the masked text (column 2) $f(x)$, the log-likelihood for the original tokens is -11.13, while the log-likelihood for the similar tokens, given the masked text, is -13.85. This example shows how when introducing the embedding of the masked text, will increase the difference in likelihood between the two candidates.

\section{Discussion}
In this paper, we investigate the performance of various embedding attack methods on sentence embedding inversion. As a comprehensive reproducibility study, our goal is to replicate the results achieved not only by the attacker model introduced in the original paper \cite{GEIA}, but also by the baseline models reported in their work. To ensure a thorough comparison, we have re-run the baseline experiments ourselves. Our findings validate that the GEIA model indeed outperforms the baselines for this task by effectively uncovering sensitive information in the input text while maintaining coherence in the inverted sentences.

Furthermore, we contribute a novel pipeline that extends GEIA, with the objective of investigating whether it is plausible for an adversary to identify sensitive information that is not part of the input text but only in the victim model's training set. After analyzing the results on a smaller subset of \cite{hidey2016identifying}, the dataset used during victim model pre-training, we have concluded that with our proposed method operating on top of GEIA, indeed, it is highly possible to recover valuable knowledge from the data the victim model was trained on, posing a serious threat for LLM security in greater scale. The latter is achieved by isolating each potential leakage from the attacker model parametric knowledge by specific pre-training and conditional access to the sentence embedding during the generation. Our study includes both quantitative and qualitative results. 

As future work, we propose masking sensitive information during training to address the training leakage observed in this study, though this may come with a potential trade-off in model performance. Evaluate this potential trade-off would give a significant understanding on sentence embeddings and their security risk.

\subsection{What was easy}
The original paper is well-written and clearly presented. Additionally, the original code is publicly available on GitHub, along with instructions for running the training scripts. Furthermore, the code for running the baseline models was also provided.

\subsection{What was difficult}
The main challenge was understanding the code due to discrepancies between the implementation and the paper. Specifically, the hyperparameter settings for the baselines differed from those reported. Additionally, the projection was not parametrized, requiring us to implement it ourselves. Furthermore, instructions for calculating the evaluation metrics were not provided. For instance, in the evaluation of PPL, a fine-tuned GPT-2 model was used, but details on how to access this model were missing. As a result of these issues, we encountered difficulties in reproducing the original results.

\subsection{Ethical and Social Impact}

In alignment with the original authors, we reaffirm our commitment to the ACM Code of Ethics and adherence to the associated code of conduct. The original work highlighted significant privacy vulnerabilities in language model-based sentence embedding models during inference, introducing generative embedding inversion as a method to further exploit these weaknesses. Building on this foundation, we extended the research to investigate potential training data leakage in language models and broadened the scope of target models to include conversational LLMs.

Our objective is to raise awareness of the privacy risks inherent in sentence embeddings and conversational LLMs while advocating for the development of robust defenses against such vulnerabilities. We emphasize the importance of conducting cohesive and systematic studies in this area, especially as these frameworks are increasingly adopted by the public, making them more susceptible to potential data breaches and exposure of personal information.

All experiments in this study were designed to ensure privacy and did not involve or expose any personally identifiable information. The datasets used, i.e., PersonaChat, QNLI and AltLex, are publicly available, with strict measures taken to maintain anonymity throughout the data collection process.

\begin{acks}
The work presented in this paper was partially supported by the LESSEN project (NWA.1389.20.183) of the research program NWA ORC 2020/21, financed by the Dutch Research Council (NWO).
\end{acks}

\appendix
\section{Prompting Schema}\label{app:promptimg_schema}

\subsection*{System/User Prompts for LLM Reasoners}
In this Appendix we report the exact prompts used for the training data leakage experiments.\\

\noindent\Large\textbf{GLM-4}\small\\
\noindent\textbf{System}: ``You are part of a team of bots with two tasks:
1. Obfuscating sensitive information in text while preserving its overall structure and meaning
2. Creating alternative versions by replacing sensitive information with similar but different entities

For each input text, you should provide TWO outputs separated by [SEP]:
- First output: Masked version using placeholders (e.g., <NAME>, <LOCATION>)
- Second output: Alternative version replacing sensitive entities with different but contextually similar entities

Rules for masking:
1. Replace sensitive entities (names, locations, organizations) with appropriate placeholders
2. Use format like <NAME>, <LOCATION>, <YEAR>, <ENTITY>
3. Maintain grammatical correctness and original meaning
4. Use consistent placeholders for identical entities
5. If no sensitive entities exist, output unchanged text

Rules for alternative version:
1. Replace sensitive entities with different but plausible alternatives
2. Maintain the same grammatical structure and coherence
3. Ensure replacements are of the same category (e.g., replace person with person, city with city)
4. The alternative should be semantically valid but change the meaning

Example 1:
Input: Barack Obama visited the United Nations headquarters in New York
Masked version: <PERSON> visited the United Nations headquarters in <LOCATION>[SEP]Alternative version: Lebron James visited the United Nations headquarters in Los Angeles

Example 2:
Input: Elon Musk is the CEO of SpaceX, based in Hawthorne, California
Masked version: <PERSON> is the CEO of <ORGANIZATION>, based in <LOCATION>[SEP]Alternative version: Sam Altman is the CEO of OpenAI, based in San Francisco

Example 3:
Input: The CEO of Tesla, Elon Musk, met with the President of the United States
Masked version: The CEO of [[ORGANIZATION]], [[PERSON]], met with the President of the United States[SEP]Alternative version: The CEO of Meta, Mark Zuckerberg, met with the Prime Minister of the United Kingdom

Please respond with: Masked version[SEP]Alternative version.''

\noindent\textbf{User}: `Please provide both masked and alternative versions for the following text: ``{prompt}'''\\

\noindent\Large\textbf{Llama-3}\small\\
\noindent\textbf{System}: You are part of a team of bots with two tasks:
1. Obfuscating sensitive information in text while preserving its overall structure and meaning
2. Creating alternative versions by replacing sensitive information with similar but different entities

For each input text, you should provide TWO outputs separated by [SEP]:
- First output: Masked version using placeholders (e.g., <NAME>, <LOCATION>)
- Second output: Alternative version replacing sensitive entities with different but contextually similar entities

Rules for masking:
1. Replace sensitive entities (names, locations, organizations) with appropriate placeholders
2. Use format like <NAME>, <LOCATION>, <YEAR>, <ENTITY>
3. Maintain grammatical correctness and original meaning
4. Use consistent placeholders for identical entities
5. If no sensitive entities exist, output unchanged text

Rules for alternative version:
1. Replace sensitive entities with different but plausible alternatives
2. Maintain the same grammatical structure and coherence
3. Ensure replacements are of the same category (e.g., replace person with person, city with city)
4. The alternative should be semantically valid but change the meaning

Example 1:
Input: Barack Obama visited the United Nations headquarters in New York
Masked version: <PERSON> visited the United Nations headquarters in <LOCATION>[SEP]Alternative version: Lebron James visited the United Nations headquarters in Los Angeles

Example 2:
Input: Elon Musk is the CEO of SpaceX, based in Hawthorne, California
Masked version: <PERSON> is the CEO of <ORGANIZATION>, based in <LOCATION>[SEP]Alternative version: Sam Altman is the CEO of OpenAI, based in San Francisco

Example 3:
Input: The CEO of Tesla, Elon Musk, met with the President of the United States
Masked version: The CEO of [[ORGANIZATION]], [[PERSON]], met with the President of the United States[SEP]Alternative version: The CEO of Meta, Mark Zuckerberg, met with the Prime Minister of the United Kingdom.

\noindent\textbf{User}: `Please provide both masked and alternative versions for the following text: ``{prompt}'''


\newpage
\bibliographystyle{ACM-Reference-Format}
\bibliography{bibliography}


\end{document}